\documentclass[final,3p]{elsarticle}
\usepackage{amsfonts}
\usepackage{amssymb}
\usepackage{amsmath}
\usepackage{graphicx}

\def\sech{\mathrm{sech}}

\usepackage[normalem]{ulem}
\usepackage{xcolor}

\newcommand{\cP}{\ensuremath{\mathcal{P}}}
\newcommand{\cT}{\ensuremath{\mathcal{T}}}

\journal{Physics Letters A}

\begin{document}

\begin{frontmatter}

\title{Solitary waves in the resonant nonlinear Schr\"odinger
  equation: stability and dynamical properties}

\author[umass]{F. Williams}

\author[umass]{F. Tsitoura}
\cortext[mycorrespondingauthor]{Corresponding author}
\ead{ftsitoura@gmail.com}

\author[uoi]{T. P. Horikis}

\author[umass,oxford]{P. G. Kevrekidis}

\address[umass]{Department of Mathematics and Statistics, University of Massachusetts Amherst,
Amherst, MA 01003-4515, USA}

\address[uoi]{Department of Mathematics, University of Ioannina, Ioannina 45110, Greece}

\address[oxford]{Mathematical Institute, University of Oxford, Oxford, UK}

\begin{abstract}
The stability and dynamical properties of the so-called resonant nonlinear
Schr\"odinger (RNLS) equation, are considered. The RNLS is a variant of the nonlinear
Schr\"odinger (NLS) equation with the addition of a perturbation used to describe wave
propagation in cold collisionless plasmas. We first examine the modulational stability
of plane waves in the RNLS model, identifying the modifications of the associated
conditions from the NLS case. We then move to the study of solitary waves with
vanishing and nonzero boundary conditions. Interestingly the RNLS, much like the usual
NLS, exhibits both dark and bright soliton solutions depending on the relative signs of
dispersion and nonlinearity. The corresponding existence, stability and dynamics of
these solutions are studied systematically in this work.
\end{abstract}

\begin{keyword}
resonant NLS equation \sep bright and dark solitons \sep stability.
\end{keyword}

\end{frontmatter}

\section{Introduction}\label{model}

Physical phenomena involving the ingredients of dispersion and nonlinearity are usually
described by nonlinear partial differential equations termed evolution
equations~\cite{waves}. A particularly interesting category of these are the so-called
integrable systems which, besides their physical significance, also exhibit remarkable
mathematical properties~\cite{ist}. Key to the study of these equations is, as one might
expect, their wave solutions.
%, which in most cases are not possible to
%find analytically.
The
Inverse Scattering Transform (IST) was developed to provide the mechanism to
systematically address such systems~\cite{ist}. When  the IST methodology is applicable,
remarkable properties can be found, such as an infinity of conserved densities,
analytical single- and multi-soliton solutions etc. Two equations stand out because they
are both physically important (in fact often thought of as {\em universal}, due to their
wide range of applications) and they initiated the field of integrable systems: the
Korteweg-de Vries (KdV) and nonlinear Schr\"odinger (NLS) models. The
former
is a prototypical model for shallow water waves while the latter is most
commonly used to describe quasi-monochromatic light propagation in
optical
media~\cite{waves}.

However, even these universal systems need to be amended in order for different phenomena
to be incorporated when considering specific mechanisms~\cite{infeld}. In this case,
non-integrable perturbations naturally emerge in these broadly applicable models, such as
the NLS equation which has been relevant to optical, atomic and water wave systems
among others~\cite{sulem,siambook}. Here, we focus on a variant of the NLS equation
which is often used to describe the transmission of uni-axial waves in a cold
collisionless plasma subject to a transverse magnetic field~\cite{lee1, lee2, lee}. This
system has quite similar features to the regular NLS equation, most notably exhibiting
bright and dark solitons depending on the relative sign of a specific parameter
(discussed below).

In its general form the resonant nonlinear Schr\"odinger (RNLS) equation \cite{lee1,
lee2, lee} reads:
\begin{eqnarray}
i \partial_t \Psi +\partial_x^2 \Psi +\gamma (-1)^{n+1}  |\Psi|^{2n} \Psi =\delta
\frac{\partial_x^2|\Psi|}{|\Psi|} \Psi,
\label{gen_RNLS}
\end{eqnarray}
where $\Psi\left(x, t\right)$ is the complex wave profile, $x$, $t$ are the spatial and
temporal variables respectively and $\gamma, \delta \in \mathbb{R}$ correspond to the
coefficients of the nonlinear terms. The last term of the equation involving
${|\Psi|_{xx}}/{|\Psi|}$, where the subscript stands for differentiation with respect to
$x$, represents
the de Broglie quantum potential, and can also be viewed as a diffraction term
\cite{Kodama}. Its coefficient, namely $\delta$, plays a crucial role in the form of the
general RNLS equation \eqref{gen_RNLS}, as it describes solutions with different
behavior, depending on the regions that are separated from the critical value $\delta =
1$ \cite{lee}. More specifically, for $\delta<1$, Eq.~\eqref{gen_RNLS} bears connections
to the NLS equation with a power law nonlinearity. For $n=1$, it admits bright and dark
soliton solutions, for $\gamma>0$ and $\gamma<0$ respectively. On the other hand, for
$\delta >1$ Eq. \eqref{gen_RNLS} has been shown to reduce to a reaction-diffusion system
(RDS) which in turn represents the simplest two-component integrable system contained in
the AKNS hierarchy of integrable systems \cite{lee,akns}. It is worth noting that similar
to the RNLS models involving nonlinear modifications of the dispersion term are of
continued interest also in other fields such as  nonlinear optics; see,
e.g.,~\cite{kurizki} for a recent (albeit somewhat different in flavor) example.

The exact solutions of Eq. \eqref{gen_RNLS} have long been discussed in the literature
and with many different methods (including the first integral method, the $G'/G$
expansion, the Darboux-B{\"a}cklund transformation and the Hirota bilinear method,
among
others)~\cite{exact1,exact2,exact3,exact4,exact5,exact6,exact7,exact8}. Here,
and
for completeness we will also derive the solitary waves
of the equation but will focus mainly on
their stability properties.
%, considering, in particular, in that regard the solitary waves of the
%model.
Below, we also discuss briefly, for completeness, some of the important properties
of the equation and refer the interested reader to Refs. \cite{lee1, lee2, lee} for more
details.

\section{Properties}

The Madelung transformation decomposes the wavefunction as $\Psi = {\rm e }^{R-iS}$ where
$R = R\left(x, t\right)$ and $S = S \left(x, t\right)$ are real-valued functions. It
follows directly that Eq.~\eqref{gen_RNLS} is equivalent to the system of equations
\begin{equation}
    \begin{gathered}
\partial_t R - \partial_x^2 S -2 \partial_x R \partial_x S = 0,   \\
%\label{MadelungR}
\partial_t S +\left(1-\delta \right) \left( \partial_x^2 R +  \left(\partial_x R\right)^2 \right)
- \left(\partial_x S\right)^2
+ \gamma (-1)^{n+1} {\rm e}^{2nR} = 0.
\label{Madelung}
   \end{gathered}
\end{equation}
First, we consider the case with $\delta<1$ and rescale time and
phase of the wave function according to
$t= \left(1- \delta\right)^{\frac{1}{2}}\tilde{t}$ and
$S = \left(1- \delta\right)^{\frac{1}{2}}\tilde{S}\left(x, \tilde{t}\right) $, respectively.
By applying the above transformations to Eq.~\eqref{Madelung}, we
get the new system for the phase and amplitude
\begin{equation}
    \begin{gathered}
\partial_{\tilde{t}} R - \partial_x^2 \tilde{S} -2 \partial_x R \partial_x \tilde{S}  = 0,   \\
\partial_{\tilde{t}} \tilde{S} + \partial_x^2 R +  \left(\partial_x R\right)^2 - \left(\partial_x
\tilde{S}\right)^2
+ \frac{\gamma}{1- \delta} {\rm e}^{2nR} = 0,
\label{Madelung2}
   \end{gathered}
\end{equation}
which, for the new complex wave function $\tilde{\Psi} = {\rm e }^{R-i \tilde{S}}$,
corresponds to the NLS equation:
\begin{eqnarray}
i \partial_{\tilde{t}} \tilde{\Psi} +\partial_x^2 \tilde{\Psi} +\frac{\gamma}{1-\delta}
|\tilde{\Psi}|^{2n} \tilde{\Psi} = 0.
\label{nls}
\end{eqnarray}
For $n=1$, the NLS equation, Eq. \eqref{nls}, for $\gamma <0$ is of the defocusing type
and describes black soliton solutions in BECs, nonlinear optics, water waves and in many
other systems. For $\gamma>0$, it admits bright soliton solutions.

However, when $\delta>1$,  the landscape of solutions
of Eq. \eqref{gen_RNLS} differs.
To illustrate this, we introduce the transformations
$t= \left(\delta -1 \right)^{\frac{1}{2}}\tilde{t}$ and
$S = \left(\delta -1 \right)^{\frac{1}{2}}\tilde{S}\left(x, \tilde{t}\right) $
and apply them to  Eq.~\eqref{Madelung}.
In particular, upon setting
\begin{eqnarray}
 r &=& {\rm exp} \left(R\left(x, \tilde{t}\right)+\tilde{S}\left(x, \tilde{t}\right)\right),   \\
 s &=& -{\rm exp} \left(R\left(x, \tilde{t}\right)-\tilde{S}\left(x, \tilde{t}\right) \right),
 \label{r_s}
 \end{eqnarray}
and by dropping the tildes, we obtain the following reaction-diffusion (RDS) system
\begin{eqnarray}
r_t -r_{xx}  + B r^{n+1} s^n&=& 0,
\label{RDSr} \\
s_t +s_{xx} - B r^n s^{n+1}&=& 0,
\label{RDSs}
\end{eqnarray}
where $B=-\gamma/\left(\delta - 1\right)$. It is particularly intriguing that from a
conservative (indeed, Hamiltonian, as is discussed below) system, we have obtained a
seemingly reaction-diffusion type system. Yet, we note that this system bears gain and
loss (notice the opposite signs in the nonlinear terms in Eqs.~(\ref{RDSr})-(\ref{RDSs}))
and can thus be brought under the umbrella of so-called $\cP \cT$-symmetric
systems~\cite{konotop}, as shown next. While we are aware of few degree-of-freedom
systems of that form~\cite{barash} (i.e., ones that could be mapped from Hamiltonian to
$\cP \cT$ symmetric or vice-versa), admittedly we were not aware of analogous
examples at the level of partial differential equations.

\subsection{Space-time invariance and Galilean symmetry of the RDS}

For the general case where $n$ is arbitrary, we investigate the time and space invariance
of the system \eqref{RDSr}-\eqref{RDSs}. If  $\tilde{r}\left(x,t\right)=
r\left(x,-t\right)$ and $\tilde{s}\left(x,t\right)= s\left(x,-t\right)$, we find that the
system is time invariant, i.e. $t \rightarrow -t$ for  $s \rightarrow r$ and $r
\rightarrow s$, i.e., the $\cP \cT$-symmetry transformation indicated above. In a
similar way, if   $\tilde{r}\left(x,t\right) = r\left(-x,t\right)$ and
$\tilde{s}\left(x,t\right)= s\left(-x,t\right)$, we reveal the space reflection of the
model, thus $x \rightarrow -x$ for $s \rightarrow s$ and $r \rightarrow r$. Next, we
investigate whether the RDS has Galilean symmetry invariance. For a fixed velocity
parameter, namely $\upsilon$, we define $a\left(x, t\right) =
\frac{\upsilon}{2}+\frac{\upsilon^2}{4}t$ and the usual Galilean transformation $x' =
x-\upsilon t $ as well as $t'=t$ can be considered. By plugging
\begin{eqnarray}
\tilde{r} \left(x, t \right) &=& {\rm e}^{-a\left(x',t'\right)} r \left(x', t'\right)
={\rm e}^{-\frac{\upsilon }{2}x+\frac{\upsilon^2}{4}t} r \left(x-\upsilon t, t \right), \\
\tilde{s} \left(x, t \right) &=& {\rm e}^{a\left(x',t'\right)} s \left(x', t'\right)
={\rm e}^{\frac{\upsilon}{2}x-\frac{\upsilon^2}{4}t} s \left(x-\upsilon t, t \right),
\label{Gal_RDS}
\end{eqnarray}
to the system \eqref{RDSr}-\eqref{RDSs} we obtain
\begin{eqnarray}
\tilde{r}_t -\tilde{r}_{xx}  + B \tilde{r}^{n+1} \tilde{s}^n&=& 0,
\\
\tilde{s}_t +\tilde{s}_{xx} - B \tilde{r}^n \tilde{s}^{n+1}&=& 0,
\end{eqnarray}
which highlights the Galilean invariance of the RDS.
{Equivalently, 
a more general solution $\tilde{\Psi}$ can be constructed by applying the Galilean
transform to $\Psi$. To do so, if  $\Psi = {\rm e}^{R-iS}$ is a solution of
\eqref{gen_RNLS}, so is $\tilde{\Psi}$ where
\begin{eqnarray}
\tilde{\Psi} \left(x, t \right) =  {\rm e}^{i \left[\frac{\upsilon}{2}x -\frac{\upsilon^2}{4}t
\right]}
\Psi\left(x- \upsilon t, t\right),
\label{psi_51}
\end{eqnarray}
for any real number $\upsilon$, ($R, S$ are real-valued again). The above follows from
the equivalence of the solutions of \eqref{gen_RNLS} with solutions of the corresponding
Madelung fluid equations, which in turn are equivalent with solutions of a corresponding
general RDS which was already proven to be Galilean invariant.}

\subsection{Integrals of motion and Lagrangian formulation}

Eq. \eqref{gen_RNLS} has also been studied with integrable systems tools (the so-called
direct methods) like the Hirota \cite{lee2} and B\"acklund-Darboux \cite{lee}
transformations. The first three conserved quantities, {for  $n=1$,} are found to be:
\begin{align}
N &= \int^{\infty}_{-\infty} |\Psi|^2 dx, \\
P &= i \int^{\infty}_{-\infty} \left(\Psi _x^{\ast} \Psi - \Psi_x \Psi^{\ast} \right) dx, \\
E &=  \int^{\infty}_{-\infty} \left( \Psi_x^{\ast}\Psi_x - \delta \left(|\Psi|_x\right)^2 -
\frac{\gamma}{2}|\Psi|^4\right) dx.
\label{integrals}
\end{align}
In different contexts these integrals represent different physical quantities. For
example, the first may correspond to the number of atoms in Bose-Einstein condensates or
to the energy of a pulse in optics. The Lagrangian of the RNLS~\eqref{gen_RNLS} {(again for  $n=1$)} reads
\begin{equation}
\mathcal{L}_{\Psi,\Psi^{\ast}} = \frac{i}{2} \left(\Psi^{\ast} \Psi_t -\Psi^{\ast}_t \Psi \right)
-
\Psi_x^{\ast}\Psi_x + \delta \left(|\Psi|_x\right)^2 + \frac{\gamma}{2}|\Psi|^4,
\label{lagr_RNLS}
\end{equation}
hence the RNLS is obtained from the Euler-Lagrange equations thereof.

Since $\Psi \left(x, t\right) = {\rm e}^{R-iS}$,
the Lagrangian~\eqref{lagr_RNLS}  can also be rewritten
\begin{eqnarray}
\mathcal{L}_{R,S} = {\rm e}^{2R} \left[ S_t-S_x^2+ \left(\delta-1 \right) R_x^2 \right] +
\frac{\gamma}{2} {\rm e}^{4R}.
\label{lagrange_Madelung}
\end{eqnarray}

We conclude our discussion in this Section by expressing $R, S$ in terms of $r, s$. Let
\begin{equation}
    \begin{gathered}
R\left(x, t\right) =  \tilde{R}\left(x, \beta t\right) = \frac{1}{2} \left[\log r \left(x,\beta
t\right) + \log \left(-s \left(x,\beta t\right) \right) \right], \\
S\left(x, t\right) = \beta \tilde{S}\left(x, \beta t\right) =  \frac{\beta}{2} \left[\log r
\left(x,\beta t\right) - \log \left(-s \left(x,\beta t\right) \right) \right],
    \end{gathered}
\end{equation}
where  $\beta = \sqrt{\delta - 1}$.
Then, the Lagrangian becomes
\begin{eqnarray}
\mathcal{L}_{r,s} = s r_t -r s_t +2r_xs_x+Br^2s^2 
\label{lagrange_RDS}
\end{eqnarray}
with the Euler-Lagrange equations coinciding with the system of Eqs.
\eqref{RDSr}-\eqref{RDSs}.

\section{Stability analysis}

A natural next step is to examine the stability of some of the simplest solutions of the
RNLS model in the form of plane wave solutions $\Psi\left(x, t \right)=A_0\exp\left[i
\left(k x-\omega t \right)\right]$, with $A_0$, $k$ and $\omega$  being the amplitude,
wavenumber and  frequency, respectively, of  Eq.~\eqref{gen_RNLS}, with  $n=1$, {by  
performing} the standard modulation instability (MI) analysis. To that effect, we consider
the stability for the most trivial case where $k=0$ and  $A_0=\sqrt{-\omega/ \gamma}$. In
that realm, we introduce the following ansatz
\begin{eqnarray}
\Psi \left(x, t\right) = \left(\Psi_0+ \epsilon b\left(x, t \right) \right) \exp \left[i
\left(-\omega t  + \epsilon w\left(x, t\right) \right) \right], \qquad 0<\epsilon\ll 1,
\label{MI_ansatz}
\end{eqnarray}
and plug it into Eq. \eqref{gen_RNLS}. The amplitude and phase perturbations assume the
form $b\left(x, t \right)=b_0 \exp \left(i \left(Q x- \Omega t\right)\right)$ and
$w\left(x, t \right)=w_0 \exp \left(i \left(Q x- \Omega t\right)\right)$ where $b_0$ and
$w_0$ are constants whereas $Q$ and $\Omega$ are the perturbation wavenumber and
frequency, respectively. At order $ \mathcal{O} \left(\epsilon \right)$, the following
dispersion relation is obtained
\begin{eqnarray}
\Omega^2 =Q^2 \left[ \left(1-\delta \right) Q^2-2 \gamma A_0^2 \right],
\label{pert_disprel}
\end{eqnarray}
that links the perturbation wavenumber, frequency and amplitude of the solution. When
$\delta=0$,
% and $n=1$,
Eq. \eqref{pert_disprel} coincides with the respective dispersion
relation of the NLS equation~\cite{sulem,siambook}. In particular, the results arising
from the stability analysis of the NLS equation  and for $\gamma=-1$ suggest the absence
of that instability  and the plane waves are spectrally stable. On the other hand, for
the focusing NLS case with $\gamma=1$, we expect that small perturbations with
wavenumbers $Q<Q_{cr}=A_0\sqrt{2 \gamma}$, on top of the plane wave solution, will grow
over time. For the RNLS equation with $\delta<1$, the MI analysis results are similar to
the ones of the NLS equation. More specifically, and for $\gamma<0$, there is no
modulation instability, and thus the plane waves persist. As such, localized solutions of
interest may occur on top of a plane wave background, whereas for $\gamma>0$,  we expect
that suitable small perturbations will grow over time for $Q<Q_{cr}=A_0\sqrt{2
\gamma/(1-\delta)}$ in that case. On the other hand, when $\delta>1$, and for any value
of $\gamma$,  small perturbations with wavenumbers $Q^2>2 A_0^2 \gamma/\left(1-\delta
\right)$  grow over time. Thus, we expect instabilities to occur for arbitrary
wavenumbers in the focusing case of $\gamma>0$; additionally, an unstable regime emerges
even for the defocusing regime of $\gamma<0$, but for small-wavelength perturbations
contrary to the $\gamma<0$, $\delta<1$ setting. The above analysis, suggests two
different regimes in the parameter $\delta$, that is,  $\delta<1$ and $\delta>1$. We will
treat the cases with $\delta<1$ and $\gamma<0$, and  $\delta>1$ (for all values of
$\gamma$), called scenarios A and B hereafter.

\begin{figure}[htbp]
\centering
\includegraphics[scale=0.239]{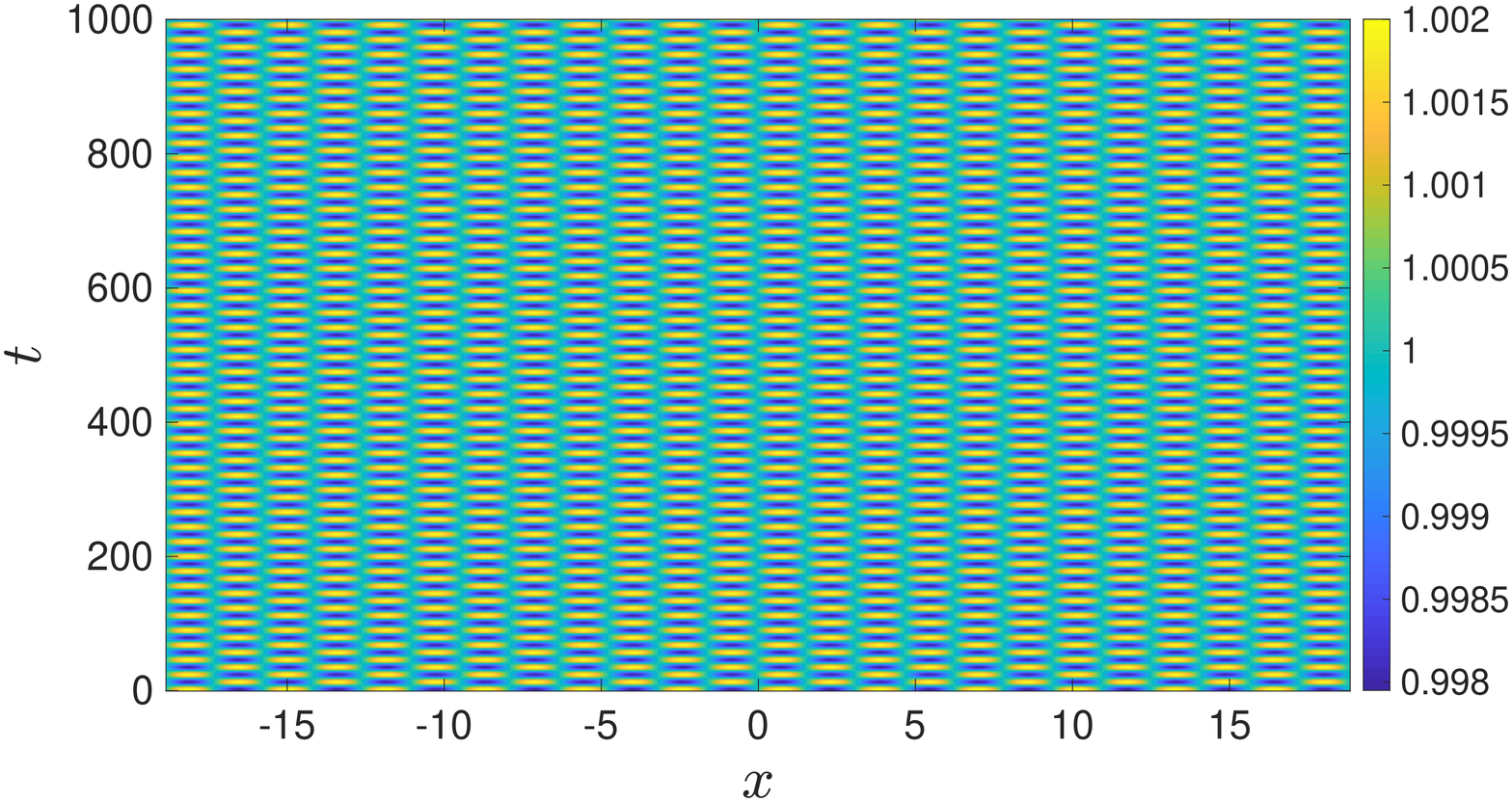}
\includegraphics[scale=0.239]{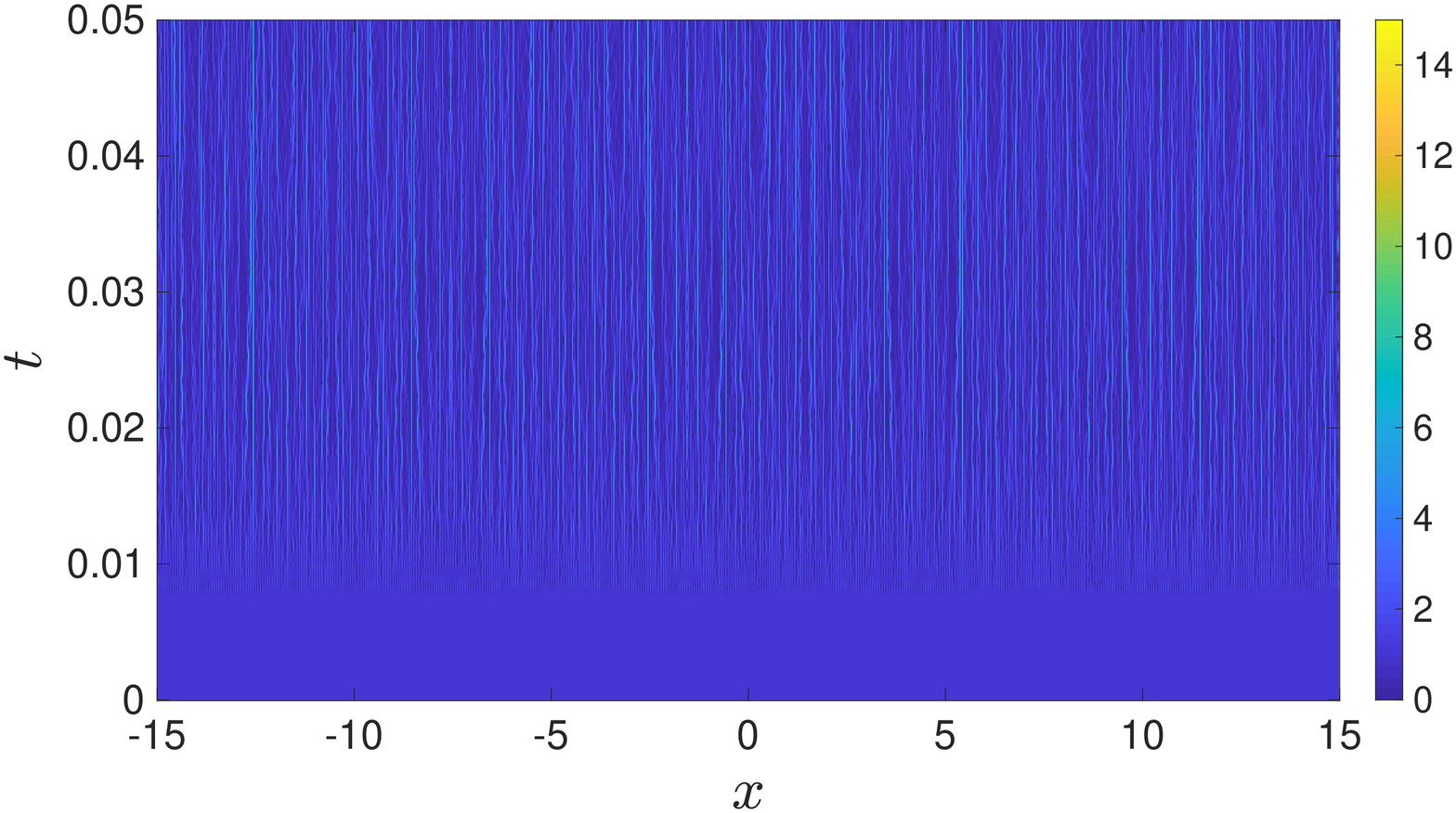}
\caption{Contour plots of plane wave evolution of the RNLS equation for
$\delta<1$ and $\delta>1$ shown in the left and right panels respectively.
The initial condition is $\Psi\left(x,t=0\right)=1$
perturbed by a sine function (see text for details).
The parameters used are  $\delta=0.99$ and $\gamma=-1$, and
$\delta=2$, $\gamma=-1$, in the left and right panels respectively.}
\label{evol}
\end{figure}

It is therefore relevant to show the dynamical evolution of plane wave solutions in those
two cases. The results confirming the MI analysis are depicted in Fig. \ref{evol} for
both scenarios A (left panel) and B (right panel), respectively. In particular, these
panels correspond to densities $|\Psi\left(x, t\right)|^{2}$ for the initial condition
$\Psi\left(x, t=0\right)=1$ perturbed by a sine function of amplitude of
$\mathcal{O}\left(10^{-3}\right)$ that has a wavenumber $Q$ of $Q = 2$. When $\delta<1$
and $\gamma=-1$ (left panel), there are no signs of instability, as predicted from
Eq.~\eqref{pert_disprel}, however when $\delta>1$ (right panel)  the plane wave solution
is highly unstable. It should be noted that we numerically integrate Eq.~\eqref{gen_RNLS}
in time using a finite difference scheme for the spatial discretization and the
Runge-Kutta method for the time marching. The resolution of the spatial discretization is
$dx=0.02$ and the time step-size is $dt=10^{-4}$ for the left panel and $dt=10^{-6}$ for
the right panel of Fig. \ref{evol}.

\subsection{Linear stability analysis of the stationary states} \label{bdg}

Once a steady state solution is available, we can explore their potential robustness.
Towards this direction, we use the linear stability analysis, and upon linearizing Eq.
\eqref{gen_RNLS} around a steady state, we find its excitation spectrum. The scope of
this analysis is to identify the fate of small perturbations on top of an already known
solution. If $\phi \left(x \right)$ stands for a steady state, we consider small
perturbations through the ansatz:
\begin{eqnarray}
\Psi\left(x, t\right) = {\rm e}^{-i\mu t} \bigg\{ \phi \left(x \right) + \epsilon \left[u
\left(x\right) {\rm e}^{-i\omega t} + \nu^{\ast} \left(x\right)  {\rm e}^{i\omega^{\ast} t}
\right]\bigg\},
\label{bdg_ansatz}
\end{eqnarray}
where $\omega$ is the (potentially) complex eigenfrequency, i.e. $\omega=\omega_r+i\omega_i$ and
$\left(u, \nu \right)$ are the perturbation eigenmodes. After plugging the ansatz
\eqref{bdg_ansatz} into Eq.~\eqref{gen_RNLS} and keeping terms of order
$\mathcal{O}\left(\epsilon \right)$, we derive the eigenvalue problem:
\begin{equation}
\left[
  \begin{array}{cccc}
  {\it L_1}        &   {\it L_2}           \\
  -{\it L^{\ast}_2}  &   {\it L^{\ast}_1}
  \end{array}
\right]
\left[
  \begin{array}{cccc}
 u         \\
 v
  \end{array}
\right]    =
\omega
\left[
  \begin{array}{cccc}
 u \\
 v
  \end{array}
\right],
\label{eig_syst}
\end{equation}
where ${\it L_1}$ and ${\it L_2}$ are given by
\begin{align}
{\it L_1} &= -  \left( 1-\frac{\delta}{2}\right) \partial_{xx}
                     +  \frac{\delta}{4}   \left(  - 2 \left(\frac{\phi_x}{\phi}
                                                            - \frac{\phi^{\ast}_x}{\phi^{\ast}}
                                                            \right)
                                                           + \left( \frac{\phi_x}{\phi} \right)^2
                                                            - \left(
                                                            \frac{\phi^{\ast}_x}{\phi^{\ast}}
                                                            \right)^2
                                                           + 2
                                                           \frac{\phi^{\ast}_{xx}}{\phi^{\ast}}
                                                           \right)
                        -\mu -\gamma \left(-1 \right)^{n+1} \left(n+1\right) |\phi|^{2n},
                        \\
{\it L_2} &= \frac{\delta}{2}  \left(  \frac{\phi}{\phi^{\ast}}  \partial_{xx}
                                                      + \left( \frac{\phi_x-\phi \phi^{\ast}_x}{\phi^{\ast}} 
                                                       \right) \partial_x
                                                       - \frac{|\phi_x|^2}{\left(\phi^{\ast}
                                                       \right)^2} + \frac{\phi \left(
                                                       \phi^{\ast}_x
                                                       \right)^2}{\left(\phi^{\ast}\right)^3}
                                                       - \frac{\phi \phi^{\ast}_{xx}
                                                       }{\left(\phi^{\ast}\right)^2}     \right)
                         -\gamma \left(-1 \right)^{n+1} n \phi^2 |\phi|^{2\left( n-1 \right)}.
\label{L}
\end{align}
For the simplest case where $n=1$ and $\phi\left(x \right) \in \mathbb{R}$, the above
matrix elements assume the much simpler form
\begin{align}
{\it L_1} &= -\left(1-\frac{\delta}{2} \right) \partial_{xx} +
\frac{\delta}{2}\frac{\phi_{xx}}{\phi}-2\gamma \phi^2-\mu, \\
{\it L_2} &=  \frac{\delta}{2}  \left( \partial_{xx} - \frac{\phi_{xx}}{\phi} \right)  -\gamma
\phi^2.
\label{bdg_n1}
\end{align}
Steady states are numerically identified by using the Newton-Raphson method, although as
we will see below they are often available in closed analytical form for the steady
states waveforms of the present model. Then their spectrum $\left(\omega_r,
\omega_i\right)$ is obtained by solving numerically the eigenvalue
problem of  Eq.
\eqref{eig_syst}. When the spectrum has purely real eigenfrequencies, then the
corresponding steady state is dynamically \textit{stable}. The presence of complex
eigenfrequencies in the spectrum indicates that the particular steady state is unstable,
i.e., that small perturbations lead either to an exponential growth (for purely
imaginary eigenfrequencies) or to an oscillatory instability of the solution (for
genuinely complex ones --although such a scenario will not be encountered in what
follows). In the sections that follow, we are going to investigate the solutions that
exist for scenarios $A$ and $B$ and monitor their stability and associated dynamics.

\section{Scenario A: Soliton solutions for  $\delta<1$ and $\gamma<0$}  \label{sec.dark}

\begin{figure}[htbp]
\centering
\includegraphics[scale=0.239]{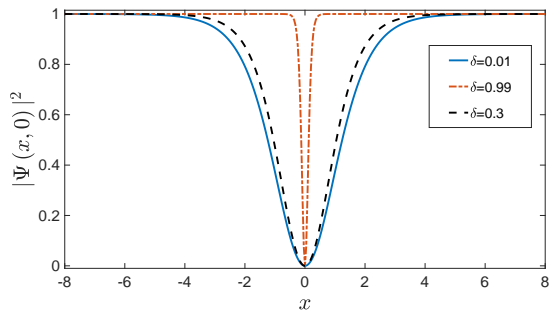}
\caption{Densities of the stationary dark solitons
  for three different values of the
parameter $\delta$ of $\delta=0.01$ (solid blue), $\delta=0.3$ (dashed black
line), and $\delta=0.99$ (dashed-dotted red line), respectively. The other
parameter values are: $\gamma=-1$ and $\mu=-1$.}
\label{dark_sols}
\end{figure}

This case scenario corresponds to $\gamma<0$ and $\delta<1$. A fundamental solution that
can be obtained in this effectively (still) defocusing case is the dark soliton solution,
and its  existence, stability and dynamics will be investigated in what follows.

%{\sout{Note, that while only the single (bright and dark) soliton solutions will be discussed
%the general cnoidal solutions of the system are also included in the Appendix for
%completeness.}}

\subsection{Exact solutions}

For the case of cubic nonlinearity ($n=1$), Eq. \eqref{gen_RNLS} possesses
dark soliton solutions of the  form
\begin{eqnarray}
\Psi \left(x, t \right) = \sqrt{-\mu}  \tanh \left(
  \sqrt{\frac{-\mu}{2 (1-\delta)}} x \right){\rm
e}^{i\mu t }, \quad \mu<0,
\label{dark}
\end{eqnarray}
where we set $\mu=-1$ in the analysis that follows. Solution \eqref{dark} could be
directly obtained from the Duffing equation
\[
-\tilde{u}_{xx}+2\tilde{u}^3 +\tilde{\mu} \tilde{u}=0
\]
with $\tilde{u}=\sqrt{\tilde{\gamma}/2}u$, $\tilde{\mu} = \mu/\left(1-\delta\right)$ and
$\tilde{\gamma}=\tilde{\mu}/\mu$, which originates from \eqref{gen_RNLS} after employing
the ansatz $\Psi \left(x, t \right) = {\rm e}^{i\mu t } u$. For $\delta=0$, Eq.
\eqref{dark} gives the well-known dark soliton solution of the defocusing NLS equation.
Fig. \ref{dark_sols} illustrates solution profiles for three different values of $\delta$
of $\delta=0.01,0.3$ and $0.99$ (see the legend therein). We observe that as we go
towards the upper limit in $\delta$, the width of the solution is decreasing.

\subsection{Numerical results}

\begin{figure}[htbp]
\centering
\includegraphics[scale=0.239]{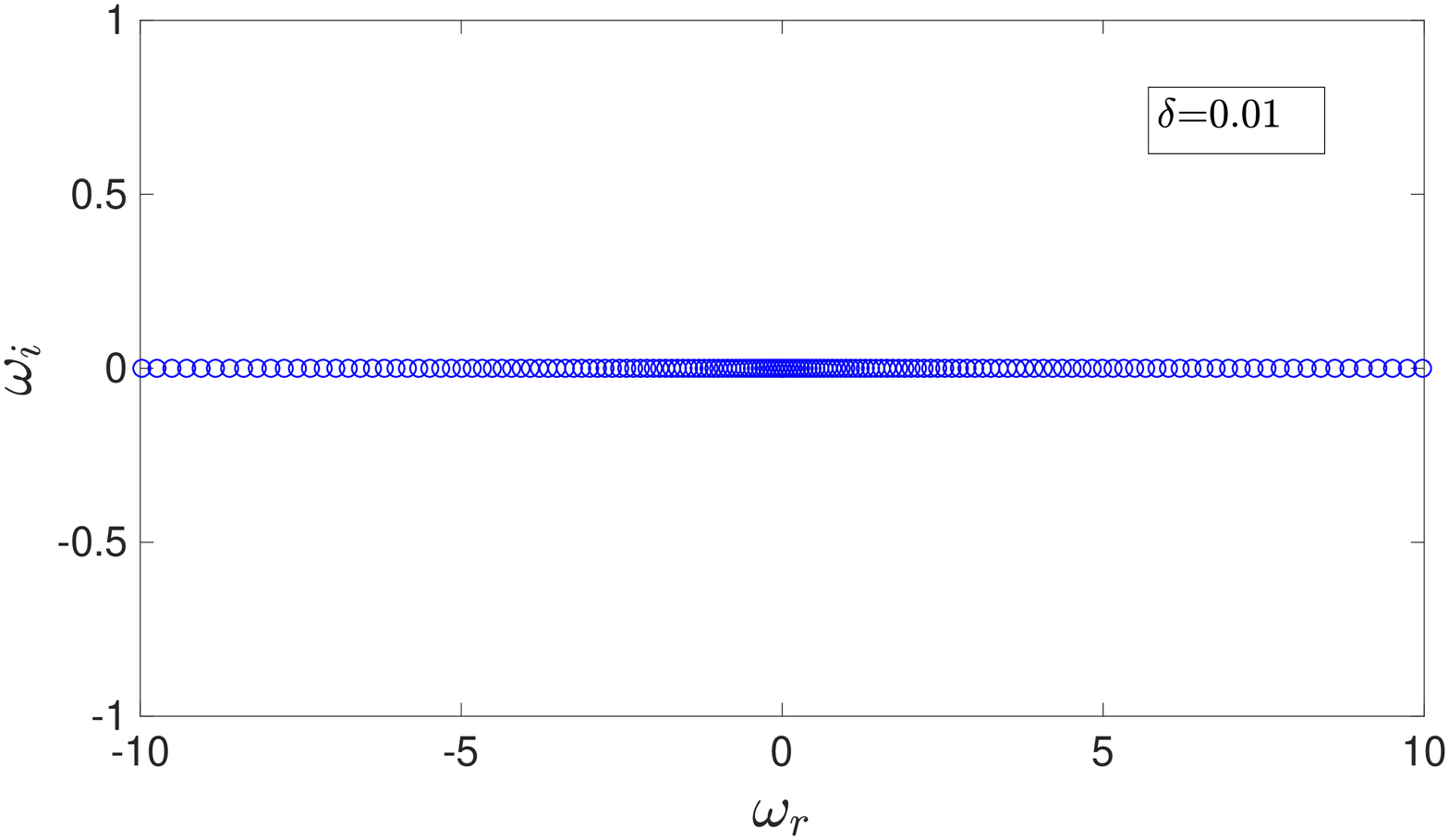}
\includegraphics[scale=0.239]{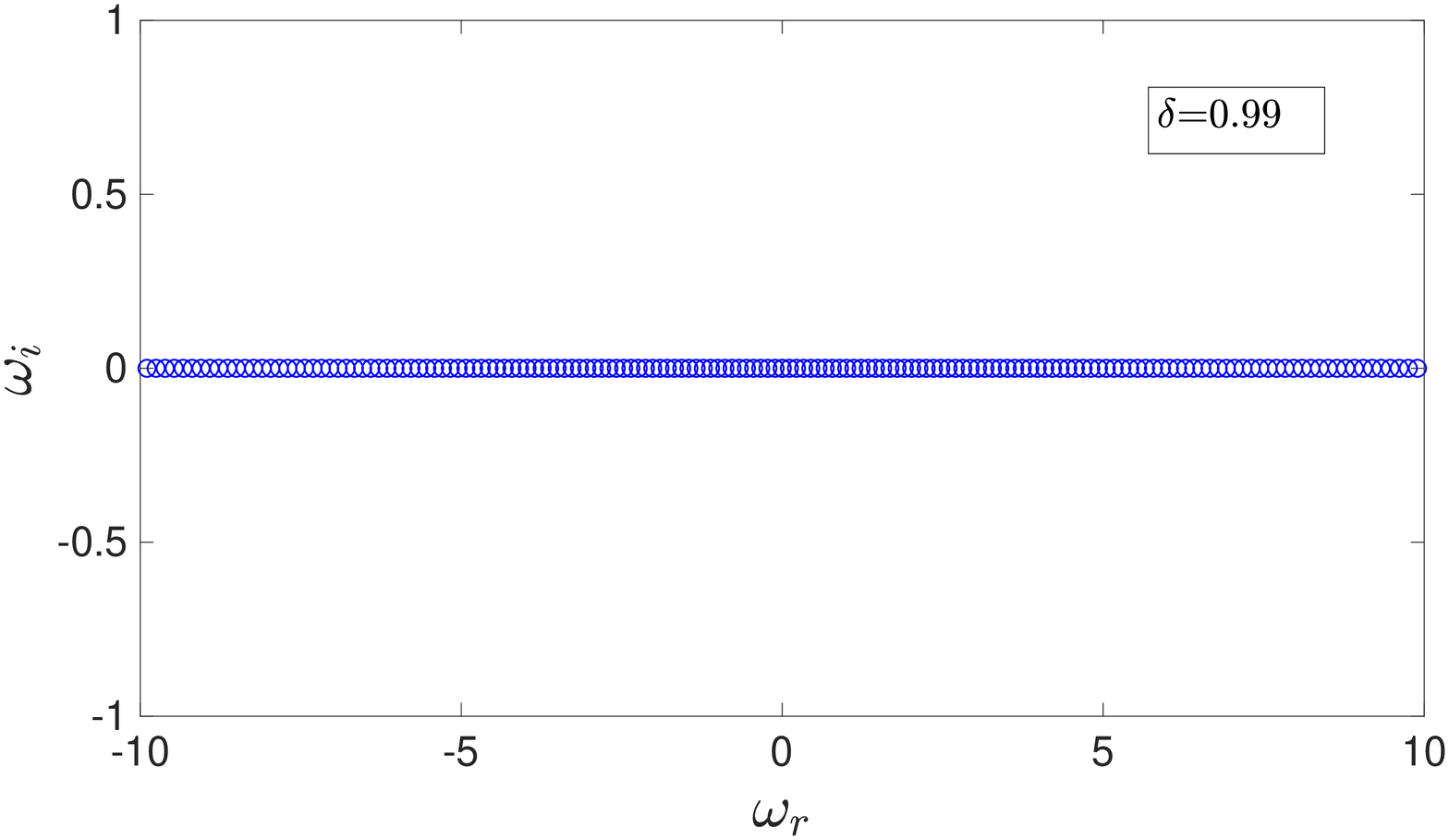}   \\
\includegraphics[scale=0.239]{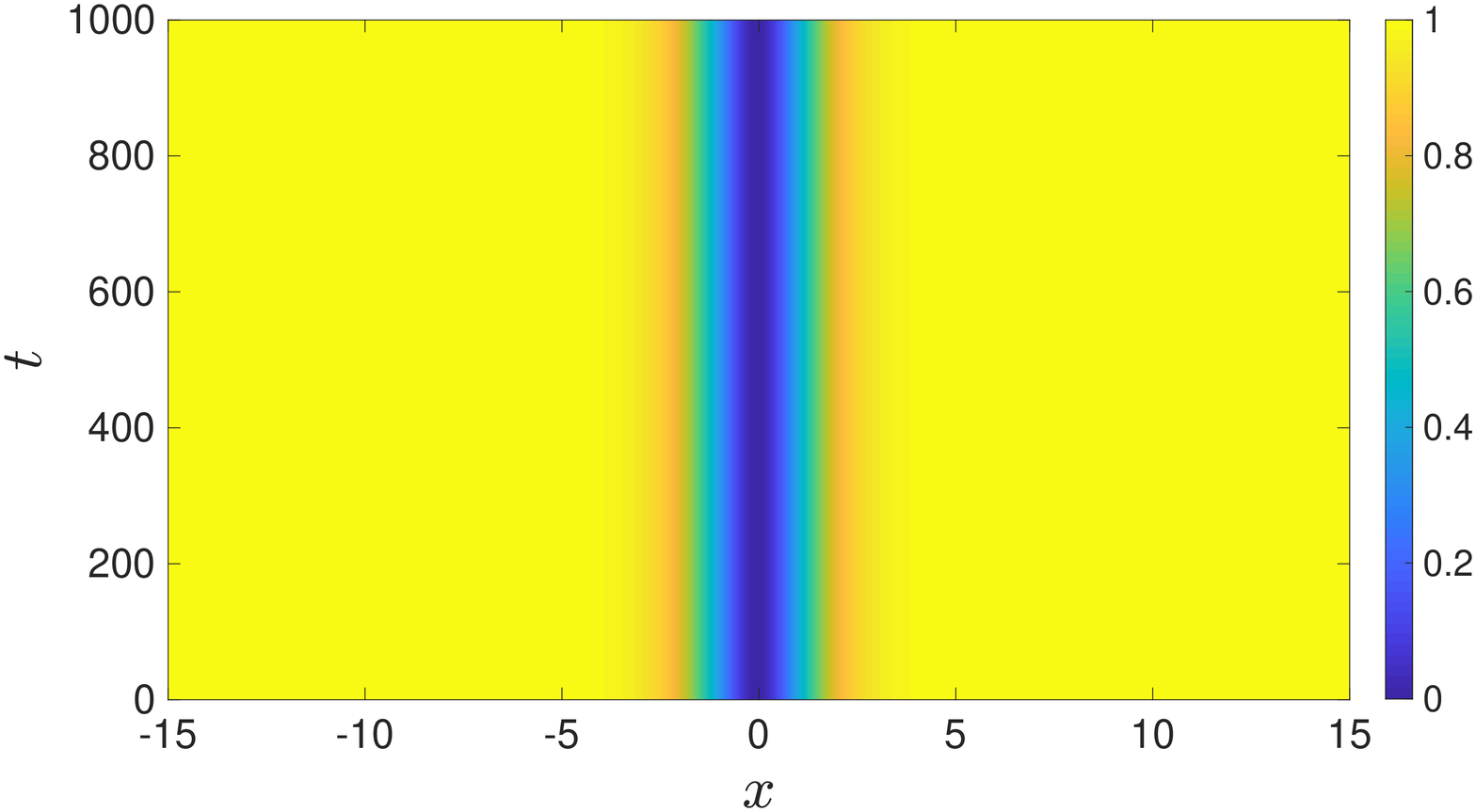}
\includegraphics[scale=0.239]{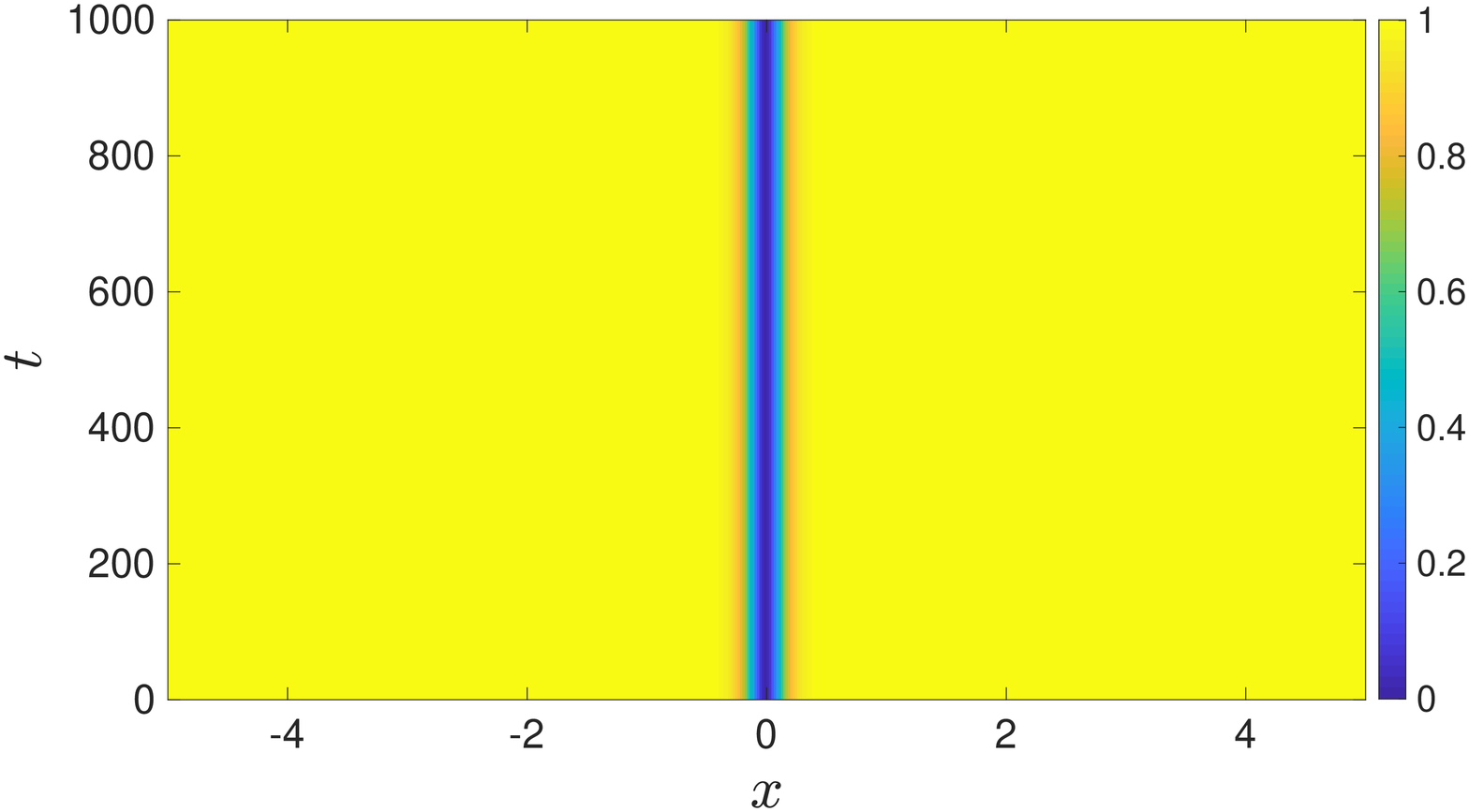}
\caption{ \textit{Top panels}: The corresponding spectral planes for the cases
$\delta=0.01$ and $0.99$ are shown in the left and right panels respectively.
\textit{Bottom panels}: Contour plots showing the dynamical evolution of the dark soliton
solutions confirming their predicted dynamical stability. The left panel shows the case
with $\delta=0.01$ and the right panel the case with $\delta=0.99$.
The other parameter values used are $\gamma=-1$ and $\mu=-1$.}
\label{dark_sp}
\end{figure}

Upon numerically identifying the above
exact solutions using the Newton-Raphson method, we proceed to
solve the eigenvalue problem \eqref{eig_syst} numerically. The corresponding spectra of
the solution \eqref{dark}  are presented in the top panels of Fig.~\ref{dark_sp} for the
cases  with $\delta=0.01$ (top left panel) and $\delta=0.99$ (top right panel),
respectively, i.e., the two limits: the NLS and the infinitesimal width case. We studied
systematically the spectra for all the values of $\delta$ in the range $\delta \in
\left(0,1\right)$ and found that all the eigenfrequencies are lying on the real axis,
thus suggesting that the solutions are spectrally stable for all $\delta$ in this
interval. It is natural to explore the robustness of  solutions \eqref{dark} by
integrating Eq.~\eqref{gen_RNLS} forward in time. To that end, we have used finite
differences for the spatial discretization with $dx=0.02$ and the Runge-Kutta method for
the time evolution with $dt=10^{-4}$. The results from the simulations are shown in the
bottom left and right panels of Fig.~\ref{dark_sp} for the cases with $\delta=0.01$ and
$\delta=0.99$, respectively. In both cases, the solitons remain robust for the time
intervals considered therein.

\section{Scenario B: Soliton solutions for $\delta>1$} \label{unstable}

\subsection{Theoretical background and exact solutions of the general RNLS equation}

We construct a traveling wave solution of the generalized RNLS \eqref{gen_RNLS} by
assuming $\Psi \left(x, t\right)= \phi \left( x-\upsilon t\right) {\rm e}^{i\left(-\kappa
x +\omega t + \theta \right)}$, where $\phi \left( w \right)>0$,  $w = x-\upsilon t$, and
$\kappa, \omega, \theta, \upsilon  \in \mathbb{R}$.

In the present case $R\left(x, t \right) = \log \phi \left(x - \upsilon t \right)$ and
$S\left(x, t \right) = \kappa x - \omega t - \theta$. Then by substituting these
transformations into Eqs.~\eqref{Madelung} with the assumption that  $\phi \left(w
\right)$ is not a constant function, we obtain the conditions:
\begin{equation}
    \begin{gathered}
\phi ''\left(w\right) + a \phi \left(w\right) = \lambda \phi \left(w\right)^{2n+1}, \\
a = \frac{\left( \omega +\kappa^2 \right)}{\delta -1}, ~~~ \lambda = - \gamma \frac{\left(
-1\right)^n }{\delta-1}.
\label{equiv_cond}
   \end{gathered}
\end{equation}
The traveling wave solution can then be found in the form $\phi \left(w\right) = A
\left[\sech \left(\alpha w\right )\right]^{1/n}$ with suitably
constants $A>0$ and $\alpha$~\cite{sulem}.
%= \left({\rm
%  const}\right)n$ \cite{sulem}.
The resulting waveform and associated algebraic conditions
for $A$, $\omega$, $\kappa$ and $\upsilon$  read:
\begin{eqnarray}
\Psi \left(x, t \right) = A \big[ \sech \left(\alpha \left(x- \upsilon t \right)\right) \big]
^{\frac{1}{n}}
  \rm{e}^{ i \left(-\kappa x +\omega t + \theta \right)},
  %\quad {\rm with}
\label{psi}
\end{eqnarray}
with
\begin{eqnarray}
\omega       = - \left( \delta-1 \right) \frac{\alpha^2}{n^2} -\kappa^2, \quad
\kappa  = -\frac{\upsilon}{2}, \quad
A          = \left[ \frac{-\left(\omega +\kappa^2 \right) \left(n+1\right)}{\gamma
\left(-1\right)^n} \right]^{\frac{1}{2n}}, \quad
\gamma \left(-1 \right)^n > 0,
\label{relations}
\end{eqnarray}
where $\theta$ is an arbitrary phase. Also, we should note  that the last condition
implies that $n$  should be odd for $\gamma<0 $, whereas $n$ should be even for
$\gamma>0$.

\begin{figure}[htbp]
\centering
\includegraphics[scale=0.239]{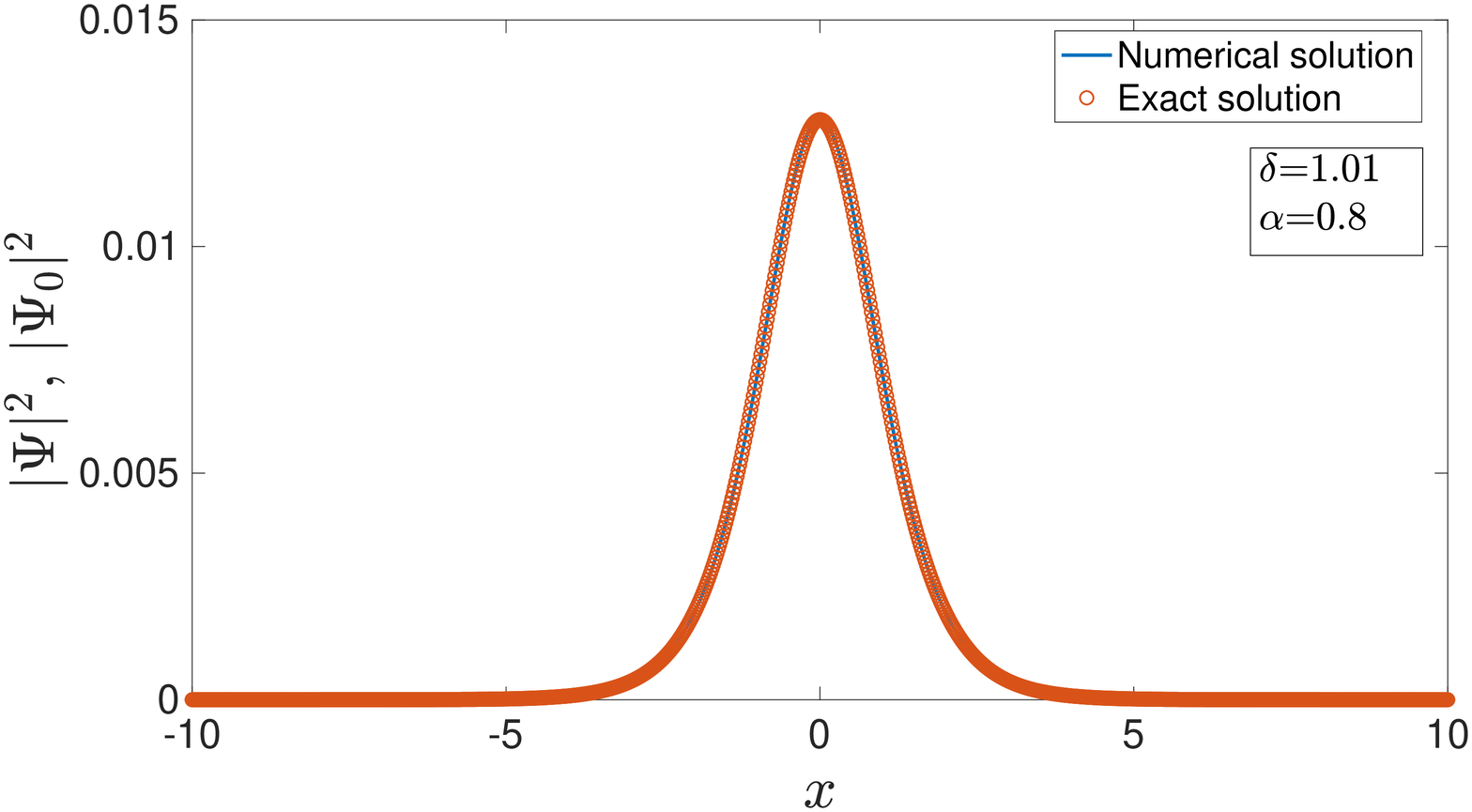}
\includegraphics[scale=0.239]{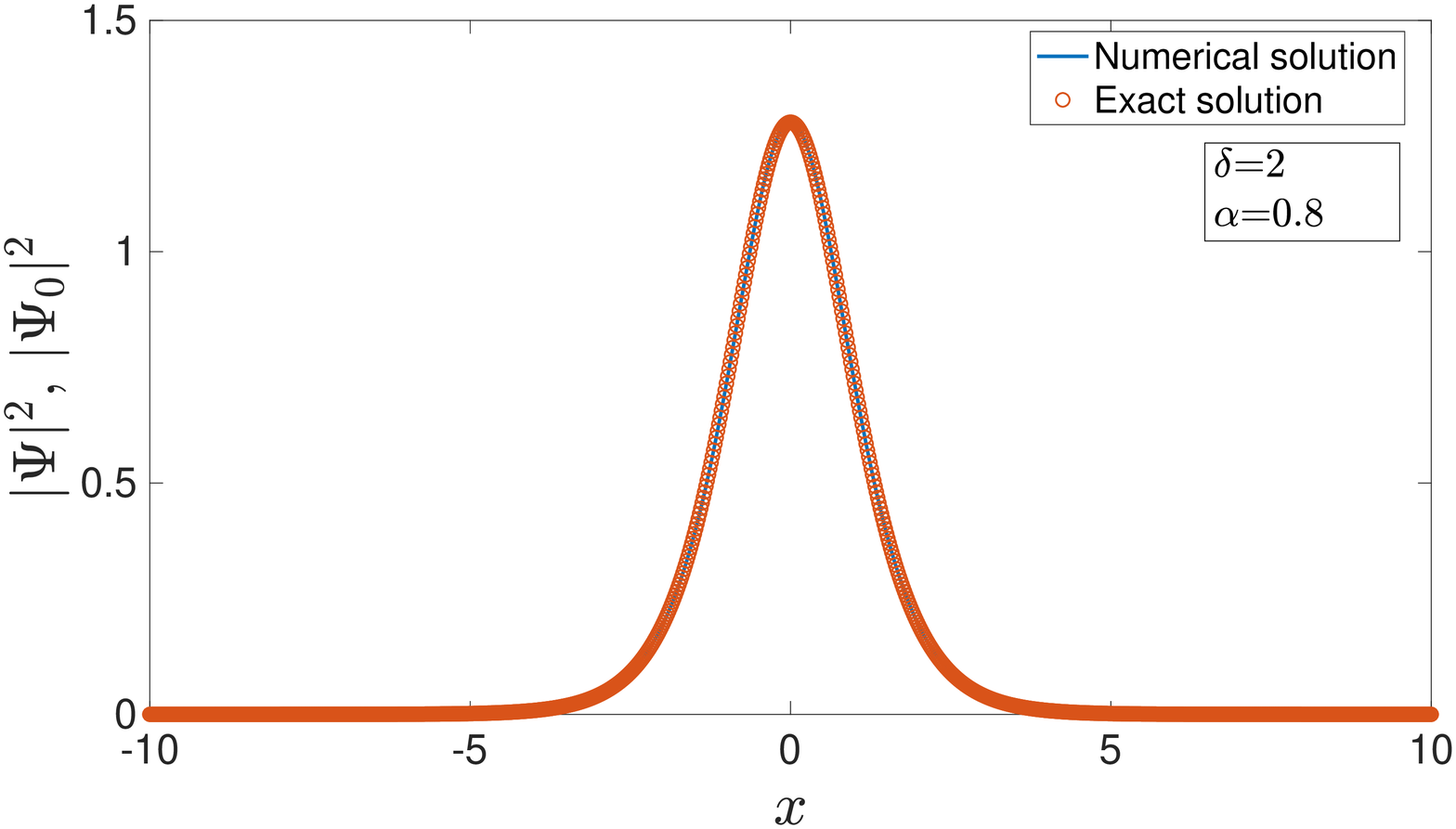}   \\
\includegraphics[scale=0.239]{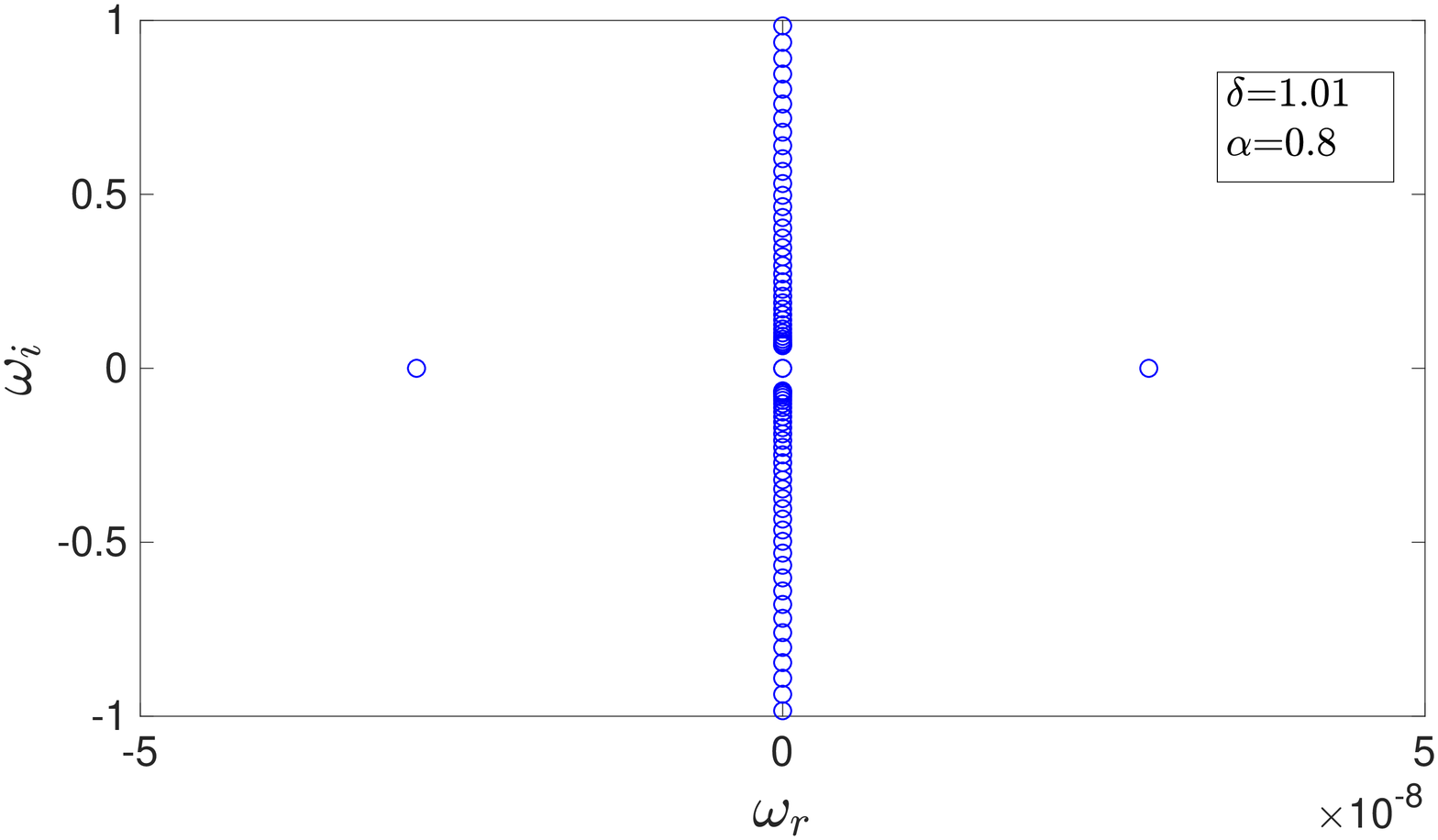}
\includegraphics[scale=0.239]{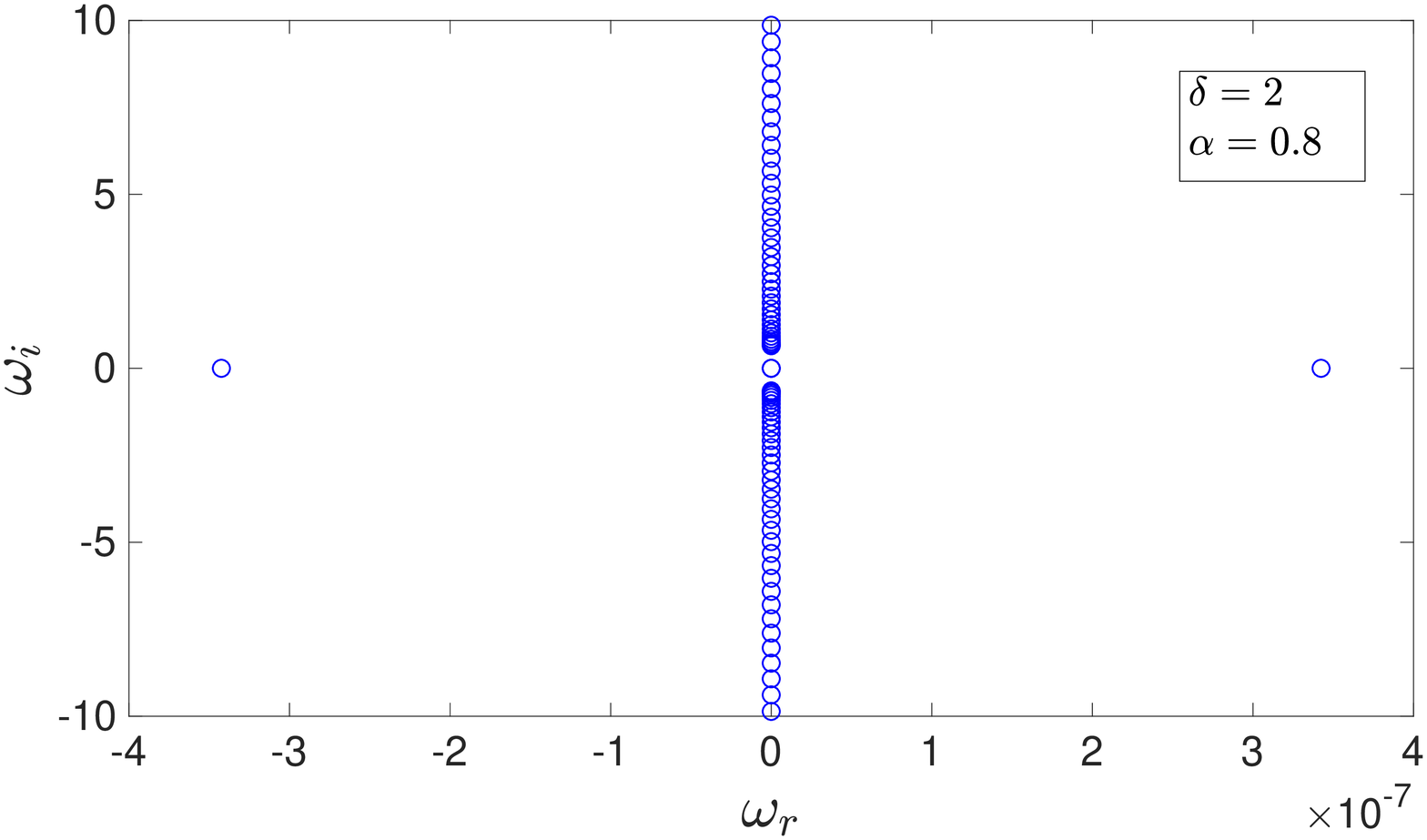}
\caption{\textit{Top panels}: Densities of the exact (red circles) and the numerical (solid blue)
steady state solutions  of Eq.~(\ref{psi2})
for $\delta=1.01$ (left panel) and $\delta=2$ (right panel).
\textit{Bottom panels}: The corresponding spectral planes.
The other parameter values used are $n=1$, $\upsilon=0$, $\gamma=-1$,
$\alpha=0.8$ and $\omega=-1$.
}
\label{fig1}
\end{figure}

\subsection{Numerical results}

We examine the relevant theoretical prediction in the prototypical case of $n=1$ for
stationary solutions with $\upsilon=0$. In that case, the associated bright solitary
waveform becomes:
\begin{eqnarray}
\Psi_0 \left(x\right)= \alpha \sqrt{\frac{2(1-\delta)}{\gamma}} \sech\left( \alpha x\right),
\label{psi2}
\end{eqnarray}
We again identify numerically the steady states of Eq.~\eqref{psi2} by using the
Newton-Raphson method. As initial guess we use the exact solution $\Psi_0$ (which
naturally constitutes an excellent initial guess, as it is exact up to the local
truncation error). We have investigated two different sets of parameters; the first
corresponds to $\delta=1.01$, $\gamma=-1$ and $\alpha=0.8$, and the second to $\delta=2$,
$\gamma=-1$ and $\alpha=0.8$. The respective profiles of the exact solutions $\Psi_0$
(red circles) and the numerically exact solutions $\Psi$ (blue line) are shown in the
top left and right panels of Fig. \ref{fig1}. The bottom panels of the same figure show
the associated spectral planes. It can be appreciated that the modes are extremely
unstable in this case. This is rather natural to expect as the point spectrum of the
solitary wave consists solely of the $\omega=0$ eigenfrequency pairs due to the
translational and phase symmetries. However, the continuous spectrum of the problem lies
entirely on the imaginary axis reflecting the modulational instability of the background,
as per Eq.~(\ref{pert_disprel}). We note in passing that rather similar results were
found for the case of $\gamma<0$ and odd $n$, as well as for that of $\gamma>0$ and even
$n$.

\begin{figure}[htbp]
\centering
\includegraphics[scale=0.27]{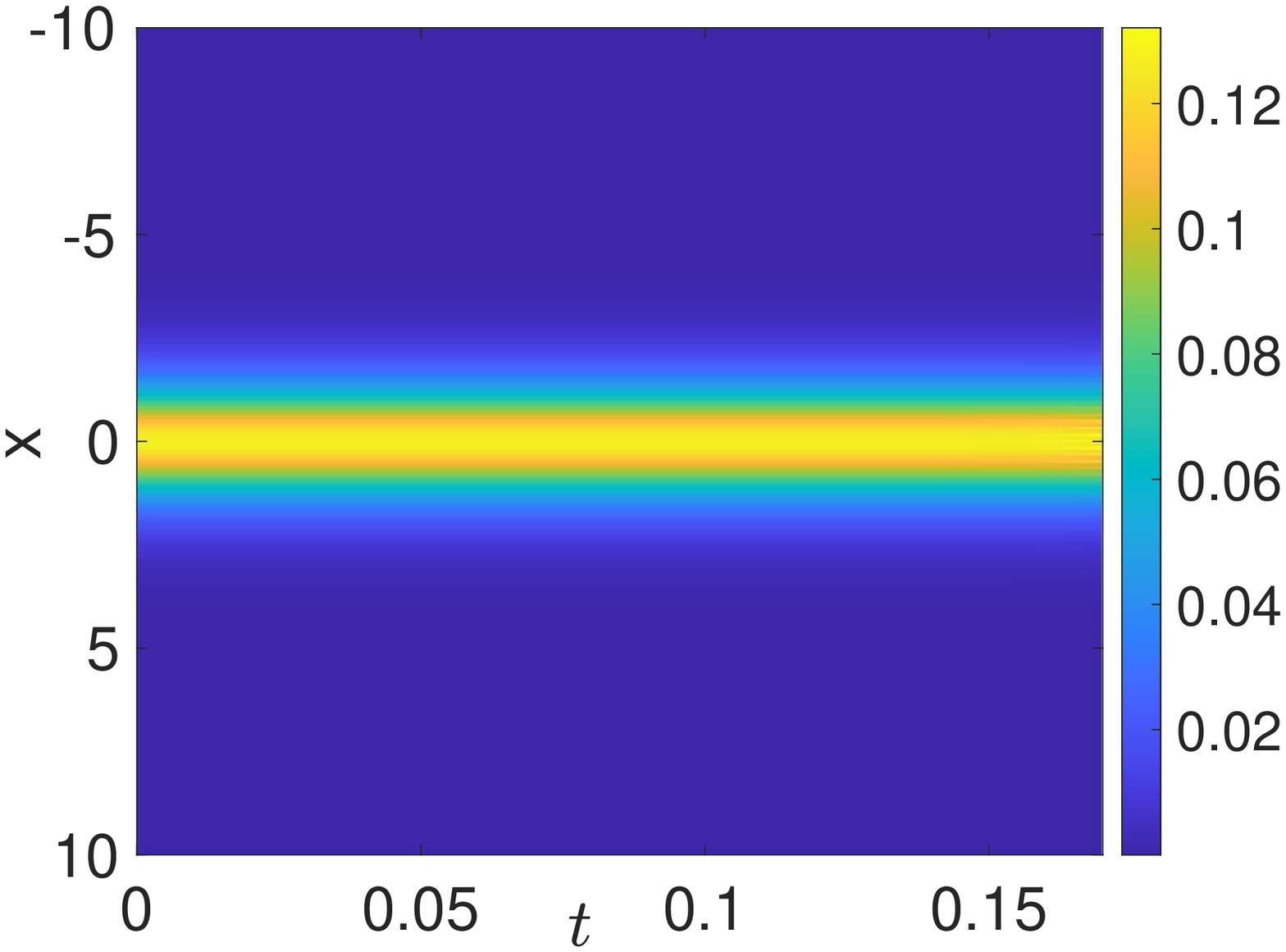}
\includegraphics[scale=0.27]{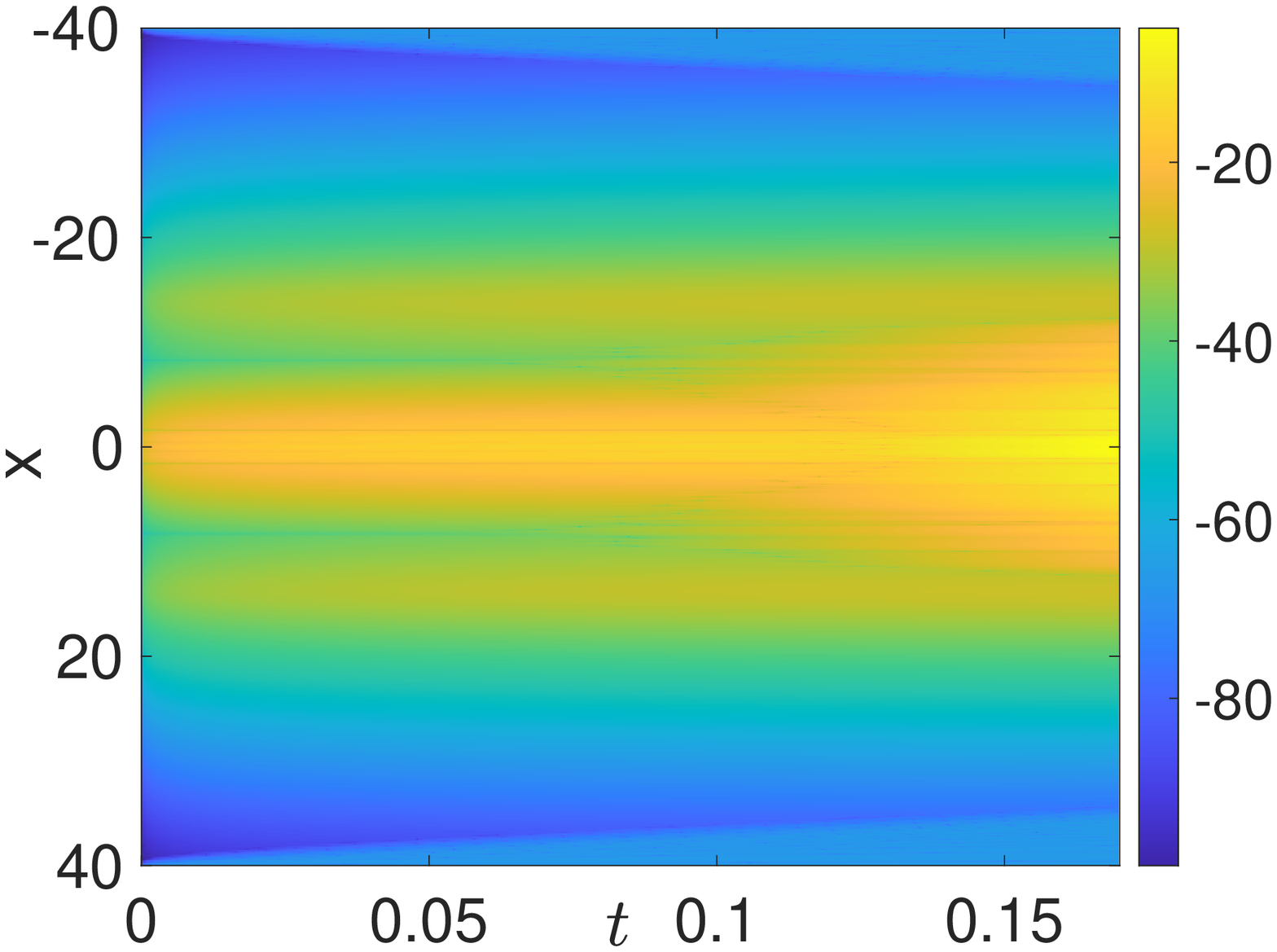}
\includegraphics[scale=0.27]{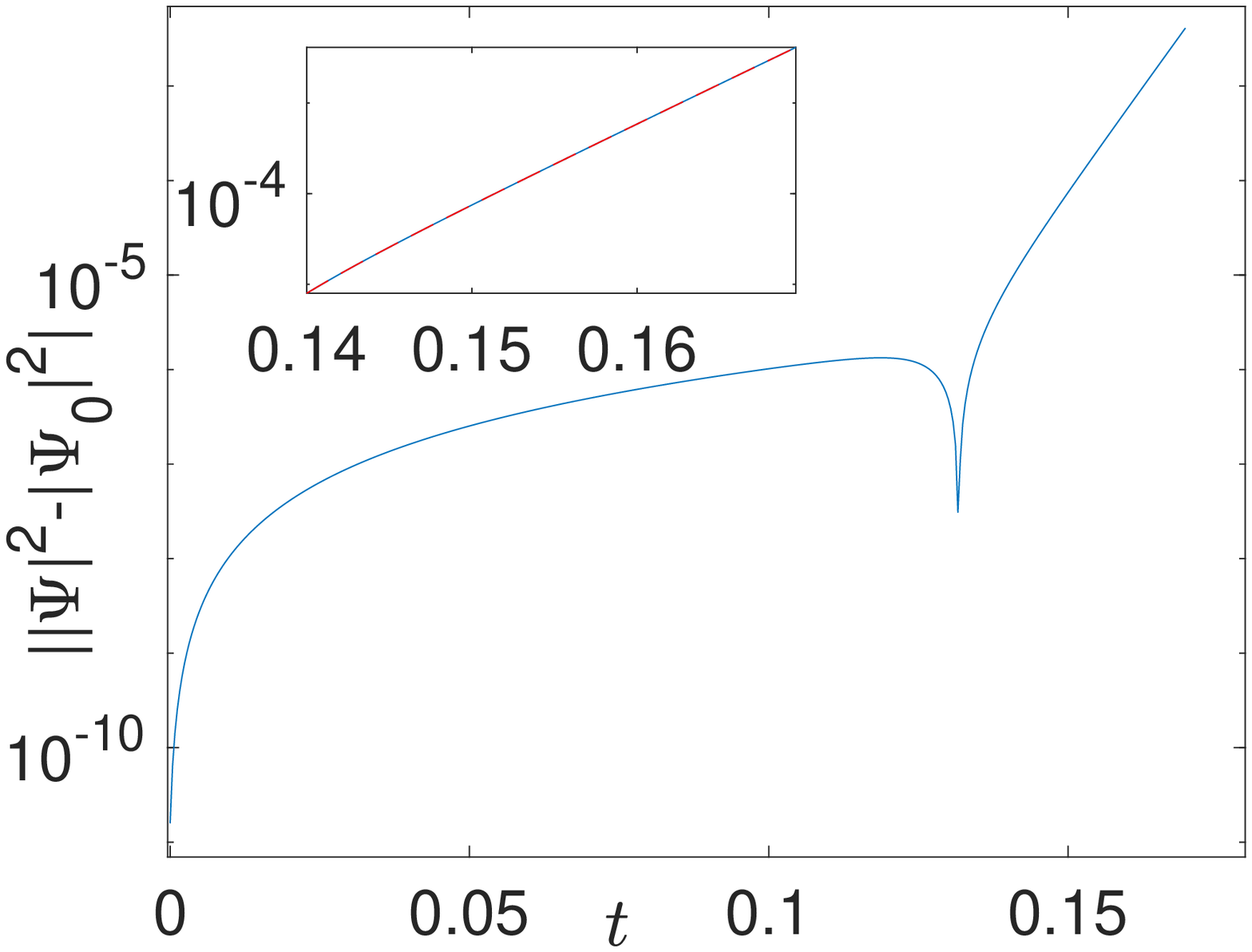}
\caption{An example of the case with $\gamma=-1$ and $\delta=1.1$ is given.
The left panel shows the space-time evolution of the norm,
where the instability is essentially indiscernible over the time
scale $[0,0.17]$ shown. The middle panel shows the (logarithm of the absolute
value of the) difference of the solution intensity from the initial condition
($\Psi_0$) intensity in its space-time evolution. Here the dramatic growth of
the error is evident (not only weakly on the boundaries but also more clearly)
in the center of the domain. Finally, in the right panel what is shown is the
quantity of the middle panel specifically at $x=0$ as a function of time. This
density difference is shown at $x=0$ as a function of time $t$ in a semilog plot.
The inset shows the exponential part of the growth with the best fit over
more than two orders of magnitude. The best fit slope (i.e., the instability
growth rate) is $208.6$.}
\label{br_sols2}
\end{figure}

Here it is important to discuss the dynamical evolution of the bright soliton solutions
for the cases with $\delta>1$  that were studied above. Notice that the stability
analysis indicates that the solution is wildly unstable with growth rates (of the
background equilibrium state) extending over an interval up to $\omega_i= Q^2
\sqrt{\delta-1}$. The maximal $Q$ corresponds to the minimal length scale $dx$ of the
domain, according to $Q=2 \pi/dx$ and is typically of the order of many hundreds. Thus,
it should come as no suprise in Fig.~\ref{br_sols2} that even when starting with the
exact analytical bright solitary wave solution with no perturbation (other than round off
error), eventually the relevant instability takes over and grows with a rate that in the
present example is found to be $208.6$. In the figure, the maximal density evolution is
shown by comparing to the density of the (exact solution) initial condition, along with
the the left panel of the space-time profile of the solution. The
latter shows no discernible signs of
instability over the time scale shown. The middle panel shows the space-time evolution
of the difference $\log(||\Psi|^2-|\Psi_0|^2|)$. The extremely rapid exponential nature
of the growth (with this extremely high growth rate) is rather transparent within the
middle and right panels of the figure, in line with our theoretical prediction about the
highly unstable nature of the $\delta>1$ dynamics.

\section{Conclusions}

We have investigated a more general form of the RNLS equation with power-law
nonlinearity. This equation, similar to the RNLS with a cubic nonlinearity, behaves
differently for different values of the coefficient of the de Broglie potential, relevant
to the cold collisionless plasma system of interest here. More specifically, for
$\delta<1$, it reduces to the standard NLS equation with a power law nonlinearity via a
suitable transformation. We investigated dark solitons for different values of the
parameter $\delta$ and we observed that they are dynamically stable solutions as our
linear stability analysis suggests and numerical simulations corroborate. On the other
hand, when $\delta>1$, the general RNLS reduces into a reaction-diffusion system which
represents a partial differential equation example of Hamiltonian form that can be
transformed into a $\cP \cT$ symmetric one. We performed a stability analysis of
bright soliton solutions of the RNLS not only with  $n=1$ but also for $n > 1$ in the
case of $\delta>1$. The analysis of these states however indicated that they are all
unstable at the linearized level, a feature in line with the modulational instability
analysis also performed herein. This instability was shown to be
associated with
large
growth rates evidenced in our direct numerical simulations.

Numerous further directions may be worthwhile to further pursue in this setting. Most
notably deriving and examining the relevant model in higher dimensions. In its one
dimensional installment and for the associated real solutions, the principal solutions of
the model appear to be similar to the usual NLS ones with suitable renormalization of the
dispersion coefficient. However, in higher dimension the genuinely complex nature of the
associated wavefunction is more likely to make it relevant for this to no longer be the
case, when considering solutions of a vortical form. There, the full role of the effect
of the $\delta$-induced perturbation may be quite intriguing to appreciate. Such topics
are currently under consideration and will be reported in future studies.

%\section*{Acknowledgements}
%% **********************************************************************
%This material is based upon work supported by the US National Science
%Foundation under Grants No. PHY-1602994 and DMS-1809074
%(PGK). PGK also acknowledges support from the Leverhulme Trust via a
%Visiting Fellowship and thanks the Mathematical Institute of the University
%of Oxford for its hospitality during part of this work.
%
%\vspace{0.4cm}

%\section*{References}

\bibliographystyle{elsarticle-num}

\end{document}